\documentclass[12pt]{article}
\usepackage{amsmath}
\usepackage{colordvi}
\usepackage{latexsym,graphicx,color,times}
\usepackage[portrait,margin=1in]{geometry}

\newcommand{\be}{\begin{equation}}
\newcommand{\ee}{\end{equation}}

\def\Alf{Alfv\'en }
\def\kapl{\chi_\parallel}

\def\bb{{\bf B}}
\def\Rdd{\Black}
\def\req#1{(\ref{eq:#1})}
\def\rfig#1{Fig.~\ref{fig:#1}}

\def\macname{hank_strauss}

\def\figdird3d{/Users/{\macname}/Documents/papers/disruption/rwtm/d3d}

\def\figdir3d{/Users/{\macname}/Documents/progs/m3dc1/d3d}

\begin{document}
\begin{center}
{\large{\bf MST Resistive Wall Tearing Mode Simulations }} \\
{H. R. Strauss$^{1, } $ \footnote{Author to whom correspondence 
should be addressed: hank@hrsfusion.com}, B. E. Chapman$^{2}$, \Rdd{N. C. Hurst$^{2}$} \\
{$^1$ HRS Fusion, West Orange, NJ 07052 } \\
{$^2$ Dept. of Physics, University of Wisconsin-Madison, Madison, WI 53706 }}
\end{center}

\abstract{
The Madison Symmetric Torus (MST) 
is a toroidal device that, when operated as a tokamak, is resistant to disruptions.  
Unlike most tokamaks, the MST plasma
is surrounded by a \Rdd{close fitting} highly conducting wall, with a
resistive wall penetration time two orders of magnitude longer than in JET or DIII-D,
and three times longer than in ITER.
The MST can operate with edge $q_a \le 2,$ unlike standard  tokamaks.
Simulations presented here indicate that the MST is unstable to resistive wall
tearing modes (RWTMs) and resistive wall modes (RWMs). They could in principle cause disruptions, 
but the predicted  thermal quench time is much
longer than the experimental pulse time. 
If the MST thermal quench time were comparable to \Rdd{ measurements in}  JET and DIII-D, 
\Rdd{theory and simulations predict that} disruptions would
have been observed in MST. 
This is consistent with the modeling herein, predicting that disruptions 
are caused by RWTMs and RWMs. 
In the low \Rdd{$q_a \sim 2$} regime of MST, 
the RWTM asymptotically satisfies the RWM dispersion relation. 
The transition from RWTM to RWM occurs smoothly at $q_a = m/n,$ where $m,n$ are
poloidal and toroidal mode numbers.}

\section{Introduction}

\Rdd{ Tokamaks are subject to disruptions, events} 
in which thermal and magnetic energy confinement  is lost.
It was not known what instability causes the thermal quench (TQ) in  disruptions. 
Disruptions have been predicted to be a severe problem in large future devices such as
ITER. %
Previous studies of JET \cite{jet21}, ITER \cite{iter21}, DIII-D \cite{d3d22}
showed that disruptions can be  caused by resistive wall tearing modes (RWTMs).
These are tearing modes (TMs) whose resonant surface is  inside the plasma, but is 
close to the wall. 
With a perfectly conducting wall, the TMs are stable, but
with no wall, they are unstable. 
\Rdd{ The TQ time is   proportional to the RWTM growth time.} 
With a highly conducting wall, the  
growth time is of order the resistive wall  penetration time. In 
the Madison Symmetric Torus (MST) \cite{mst91,mst22}
\Rdd{ the growth time } is much longer than the shot duration. 

\begin{figure}
\vspace{.5cm}
 \begin{center}
\includegraphics[width=7.5cm]{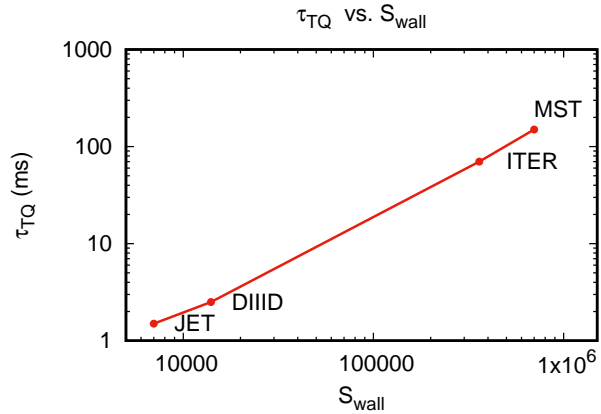}
\end{center}
\vspace{-.5cm}
\caption{\it TQ time in ms as a function of $S_{wall}$ - measured (JET, DIII-D)  and simulated (ITER, MST)}
\label{fig:ttqs1}
\end{figure}

The main result of this paper is that the predicted thermal quench time 
in MST is much longer than in conventional
tokamaks such as JET and DIII-D, and is longer than a prediction for ITER
based on RWTMs. This is shown in \rfig{ttqs1}. The TQ time in $ms$ is shown as
a function of $S_{wall}$, the resistive wall penetration time normalized
to the \Alf time (defined below). For JET and DIII-D, the TQ time is based on
experimental data and simulations. For ITER and MST, the TQ time is based on
simulation. 
The MST case is for $q_a = 2.6$, described in more detail below.
In MST,  a TQ does not occur during the experimental shot duration, 
which gives a lower limit to the possible TQ time.

Another result is that in the low edge safety factor  $q_a \le 2.6$ regime of MST, the RWTM 
growth time scales linearly in the
resistive wall penetration time \cite{d3d22}. 
{This is characteristic of large $S_{wall}$, 
in which the RWTM asymptotically satisfies the RWM dispersion relation.
ITER could also be in this regime.}
The largest amplitude  magnetic 
perturbation seen in
simulations is the RWTM with rational surface 
radius $r_s$ at which $q(r_s) = m/n$ is closest  
to $q_a,$ although RWMs can have a comparable amplitude.

\begin{figure}[h]
\vspace{.5cm}
\begin{center}
\includegraphics[height=8.0cm]{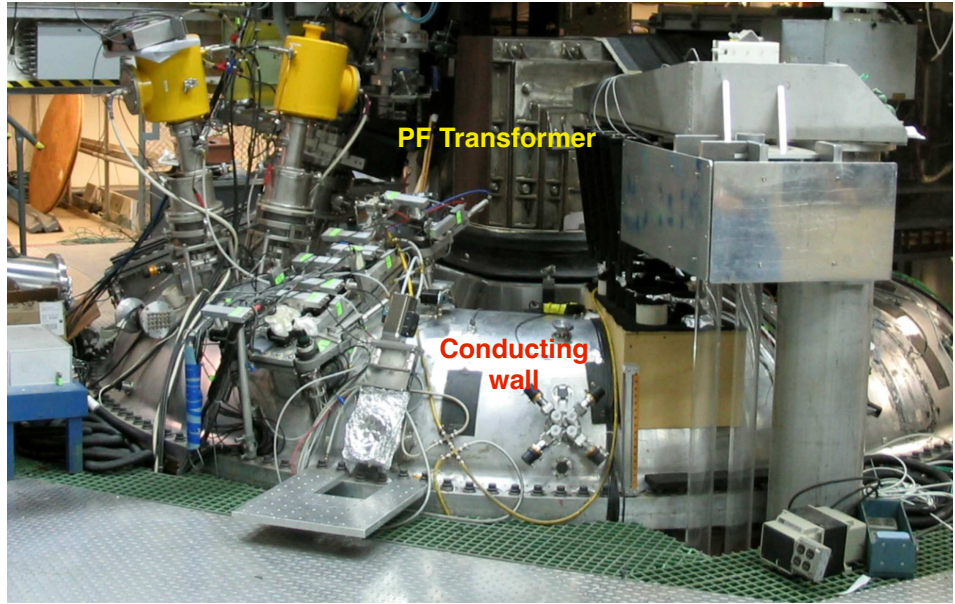}
\end{center}
\caption{\it
The MST experiment, showing the transformer and the conducting wall.
} \label{fig:mst_pic}
\end{figure}

The MST experiment 
is a toroidal device which can
be operated as a reversed field pinch (RFP) or 
as a tokamak. It is well known \cite{rfp}  that RFPs require a highly conducting wall, or
feedback, in order to stabilize external kink modes.
The MST device  is shown in \rfig{mst_pic}.
It has 
a circular cross section with limiters, 
a single-turn TF winding,
PF windings wrapped around an iron-core transformer,
and a close-fitting conducting shell with wall penetration time $\tau_{wall} =  800 ms$.

The MST wall penetration time is more than
three times longer than in ITER, with \cite{gribov} $\tau_{wall} = 250ms.$  MST has a pulse time of about $50 ms,$ during which
the wall is effectively an ideal conductor. 
It can operate with $q_a \le 2.$ \cite{mst22}.
No disruptions have been  seen to date  when it is  operated as a standard tokamak.
Disruptions can occur under non standard conditions with very low density,
in discharges dominated by runaway electrons (RE). 
It has internal MHD modes, 
including $(1,1)$ internal kinks, which produce sawteeth.
\Rdd{According to the theory and simulations presented here,}
disruptions are suppressed by the highly conducting wall.  

In the following, theory and simulations are presented which indicate that MST 
is unstable to RWTMs and RWMs, which could
cause disruptions in conventional tokamaks \cite{jet21,iter21,d3d22}. 
These results suggest that disruptions 
are not observed in MST because \Rdd{ the predicted } thermal quench time $\tau_{TQ}$ is much longer
than the experimental pulse duration. It is shown that RWTMs and RWMs  in MST
have the same mode growth time  scaling linearly in the resistive wall magnetic
penetration time $\tau_{wall}.$ It is also shown that there is a smooth transition
from RWTMs to RWMs when the rational surface exits the plasma. 

Simulations were done to obtain  the
scaling of the TQ time $\tau_{TQ}$  with  wall resistivity. The simulations were
performed with the nonlinear resistive MHD M3D code \cite{m3d} with a
resistive wall \cite{pletzer}.
Simulations were
initialized with MSTFit \cite{misfit}  equilibria having on axis  $q_0 = 1,$
and several values of edge $q = q_a$ in the range  $1.5 \le q_a \le  2.6.$ 
The parameters were:  Lundquist number  $S= 10^5$ (the experimental
value), and parallel thermal conductivity 
$\kapl = 10 R^2 / \tau_A,$ (somewhat larger than the  experimental value
 $4 R^2 / \tau_A.$) 
The experiment has $B_T = 0.133 T,$ and average density $n_e = 4.5\times 10^{12} cm^{-3}.$
The major radius is $R = 1.5m,$ and the minor radius is $0.52m.$
The fill gas was deuterium. \Rdd{The
 \Alf time is }  $\tau_A = R / v_A = 1.15 \times 10^{-6} s.$
The resistive wall time is $\tau_{wall} = \mu_0 \delta_{w} r_w /
\eta_{w},$ where $\delta_{w}$ is the wall thickness, $r_{w}$ is the wall radius,
and $\eta_w$ is the wall resistivity.
With $\tau_{wall} = .8s,$ then $S_{wall} = \tau_{wall} / \tau_A = 7\times 10^5.$  
The value of $S = 10^5,$
which gives $T = 69 eV$ on axis. The parallel conductivity in the collisional regime is
$ \kapl = 2.1 v_{te}^2 \tau_e, $ 
or $\chi_\parallel = 8 \times 10^{11} cm^2 / s.$
This can be expressed as
$ \kapl = 4.1 R^2 / \tau_A.$
\Rdd{The simulations do not include rotation, which has a stabilizing effect 
\cite{gimblett,bondeson,betti}  on RWTMs.}
\Rdd{The computational
plasma extended to the wall, and the 
 narrow limiters were not explicitly taken into account.} 


\section{RWTM Theory } \label{sec:theory}
This section reviews some   theory of RWTMs and RWMs.
The RWTM dispersion relation \cite{jet21,finn95} is
\be S^{3/4}S_{wall}^{-5/4} ( \hat{\gamma}^{9/4} + g_s \hat{\gamma}^{5/4})
= \Delta_i \hat{\gamma} + g_s \Delta_n  \label{eq:disp}  \ee
where $\hat{\gamma}  = \gamma \tau_{wall},$ $S$ is the Lundquist number, 
$g_s = 2m/[1 - (r_s/r_w)^{2m}],$
ideal wall stability parameter $\Delta_{i} = r_s \Delta_{ideal}'/m,$
external stability parameter $\Delta_{x} = 2 r_s^{2m} / (r_w^{2m}  - r_s^{2m})$,
with rational surface radius $r_s$,
in a cylindrical geometry large aspect ratio model.
The no wall stability parameter is
 $\Delta_{n} = \Delta_i + \Delta_x.$
Resistive wall tearing modes have $\Delta_{i} \le 0,$
and require finite $S_{wall}.$
The RWTM growth rate scalings vary as $\gamma \propto S^{-\alpha},$
with $4/9 \le \alpha \le 1.$ In a JET example \cite{jet21}, $\alpha = 4/9,$ 
while in a DIII-D example \cite{d3d22} $\alpha = 2/3.$

In the following MST examples, $\gamma \sim \tau_{wall}^{-1}$,  $\alpha = 1.$
This is because of the smallness of the coefficient on the 
left side of \req{disp}. With MST parameters, $S^{3/4}S_{wall}^{-5/4} = 2.8\times 10^{-4}.$
The left side of \req{disp} is only significant when $\Delta_i$ is within
a small range of $\Delta_i  = 0.$


In a step current model \cite{finn95,frs73} with 
a constant current density and $q = q_0$ contained within radius $r_0$,
zero current density for $r > r_0,$
$q = q_0 (r/r_0)^2,$ Then  $q_w = q_s (r_w/r_s)^2,$ 
where $q_0$ is the value on axis and $q_s = m/n.$
\Rdd{In the model $q_a = q_w$.}
\be \Delta_{i} = -2   \frac{n q_0 - m + 1 - (r_0/r_w)^{2m}}
{[n q_0 - m + 1 - (r_0/r_s)^{2m}][1 - (r_s/r_w)^{2m}]}. \label{eq:delta1} \ee
It can be seen that as $r_s \rightarrow r_w,$ that $\Delta_x, -\Delta_i 
\approx 2/[1 - (r_s/r_w)^{2m}].$

The growth rate is,  neglecting the left side of \req{disp},
\be {\gamma}\tau_{wall} = - 2m \frac{n q_0 - (m-1)}{nq_0 - (m-1) - (r_0/r_w)^{2m}}. \label{eq:rwm} \ee
Remarkably, this is also the growth rate of a RWM \cite{finn95,liu} using the same
model equilibrium.  The crossover from RWTM to RWM occurs smoothly
at $q_a  = m/n.$ For $q_a < m/n,$ the mode becomes a RWM.

\begin{figure}[h]
\vspace{.5cm}
\begin{center}
\includegraphics[width=7.0cm]{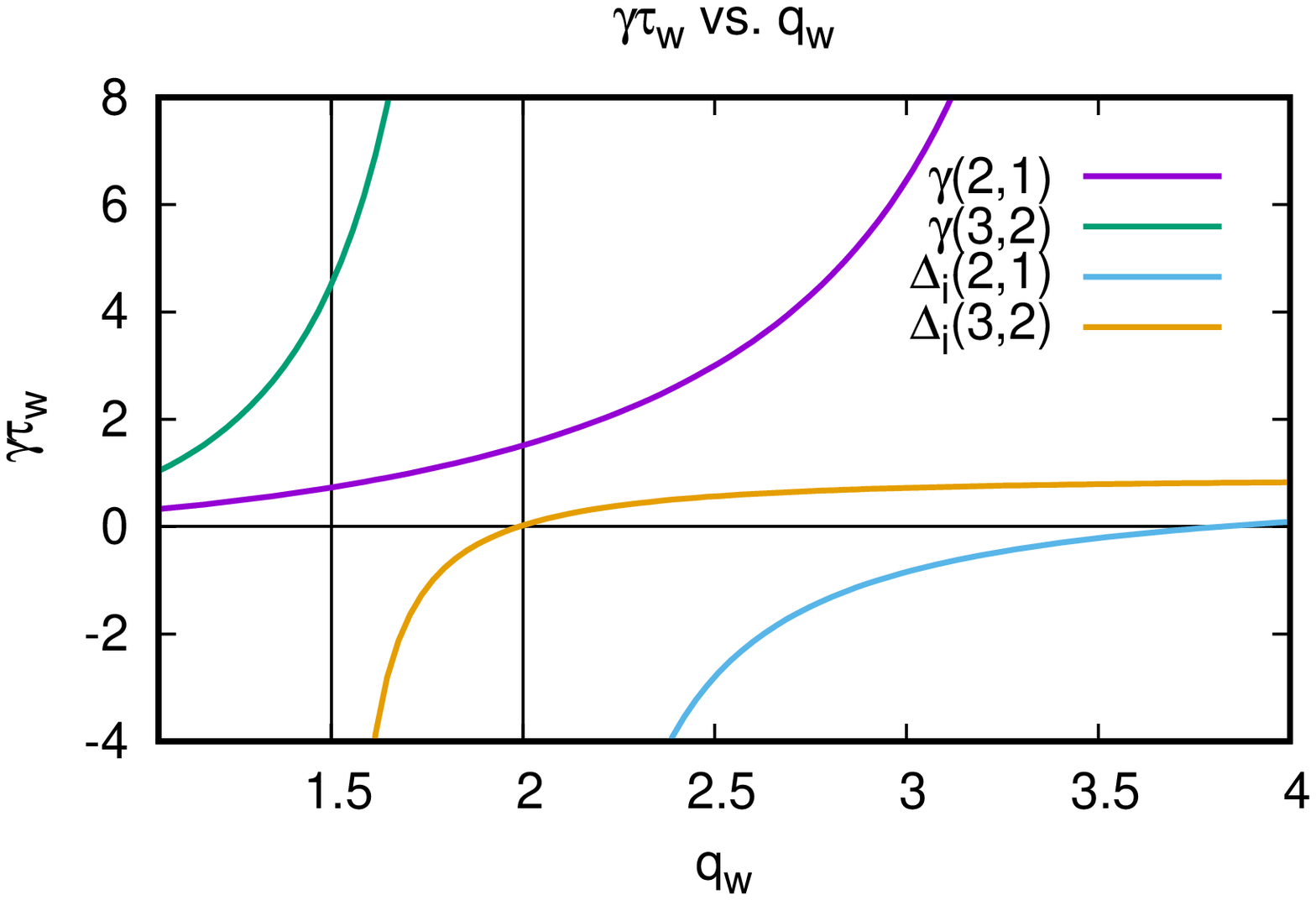}(a)
\includegraphics[width=7.0cm]{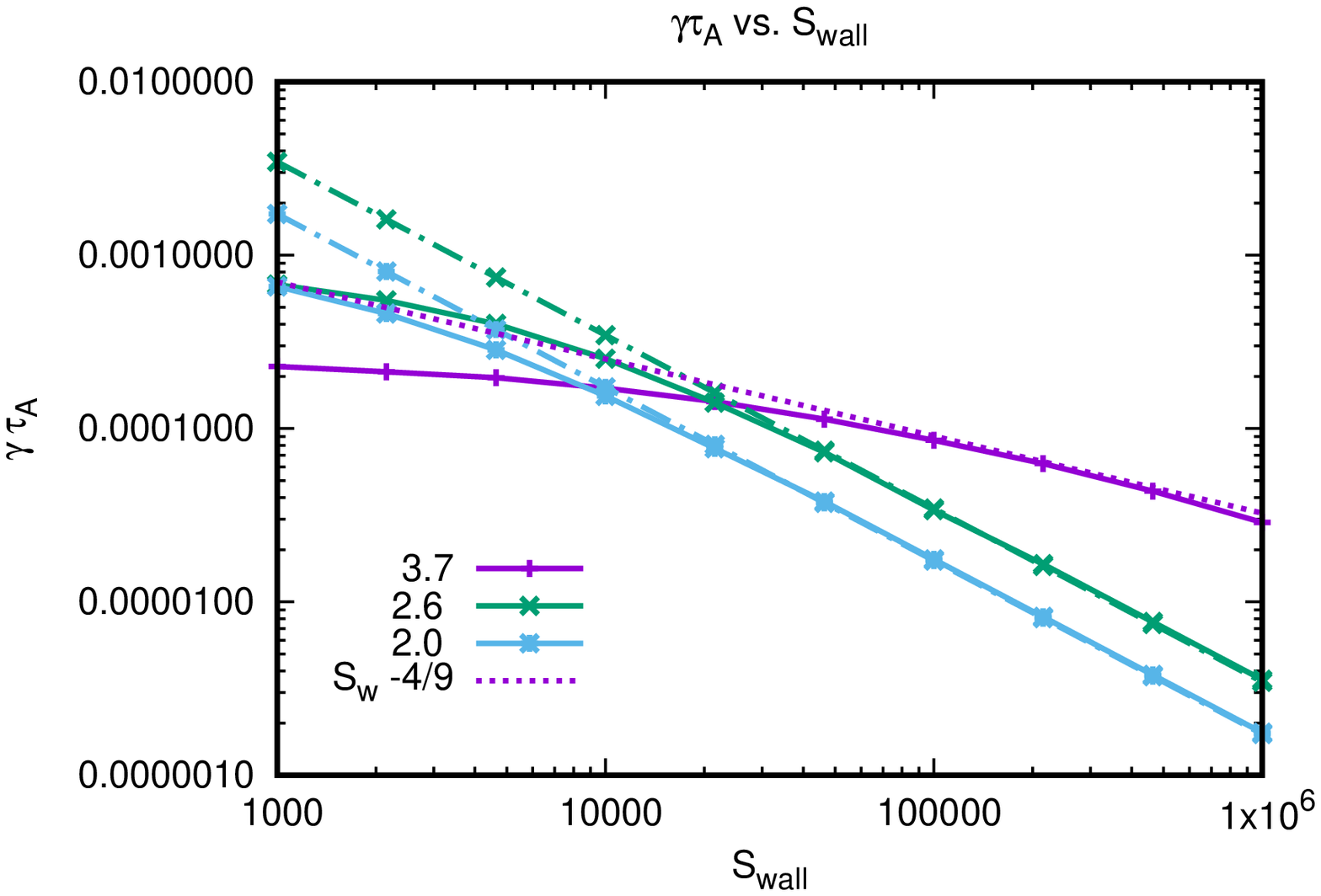}(b)
\end{center}
\caption{\it
(a) Linear growth rate $\gamma\tau_{wall}$ as a function of $q_w,$ 
\Rdd{for $(2,1)$ and $(3,2)$ modes. Also shown is $\Delta_i$ as a function of $q_w$  
for the two modes. When
$\Delta_i < 0$ the mode is a RWTM or RWM.} 
b) 
\Rdd{Growth rate $\gamma\tau_A$  as a function of $S_{wall}.$  The curves 
are labelled by the value of $q_w.$
The growth rate  asymptotes to a RWM scaling for
$S_{wall} \gg S^{3/5},$ except for $q_w = 3.7$ which asymptotes to $S_{wall}^{-4/9}.$ } 
} \label{fig:deltai}
\end{figure}

In \rfig{deltai}(a) is plotted $\gamma \tau_{wall}$ in  \req{rwm} for  the cases $q_0 = 1.08,$ $(m,n) = (2,1),$
and $(3,2)$ as a function of $q_w.$ 
Also shown is $\Delta_i$ in \req{delta1}.
For $\Delta_i \ge 0,$ the mode is a tearing mode. For $(2,1),$ this occurs at $q_w \ge 3.7,$ in
this model. 
There is a $(3,2)$ tearing mode
for $q_w \ge 2.27.$ 
For the $(2,1)$ mode, $\gamma$ is in the
regime of validity of \req{rwm} %
for $q_a \le 1.98.$ The $(3,2)$ is in this
regime for $q_a \le 1.7.$ 
There is a pole in $\Delta_i$ when $r_s = r_w$ or $q_w = m/n.$ This is the transition from
RWTM to RWM. There is no pole in $\gamma,$ so the transition from RWTM to RWM is smooth. 
There is also a pole in $\gamma$ at which 
the denominator of \req{rwm} vanishes. This is the same as the zero of $\Delta_i,$ 
at which the RWTM becomes a tearing mode.

\Rdd{The scaling of $\gamma$ with $S_{wall}$ is shown in \rfig{deltai}(b).
The curves were obtained by solving \req{disp} with parameters
$(m,n) = (2,1)$, $q_0 = 1.08,$ $S = 10^5,$ and several $q_w$ values.
For small $S_{wall},$ the RWTM satisfies a no wall tearing mode dispersion relation.
For large $S_{wall},$
except for the case $\Delta_i = 0,$ with $q_w = 3.7,$ which has
$\alpha = 4/9,$ asymptotically 
$\gamma $ is given by \req{rwm}, plotted as  dashed curves. The range of
$S_{wall}$ which deviates from \req{rwm} decreases as $q_w$ decreases.  }

The model used here is 
consistent  with simulations.
The RWTM and RWM have a growth time and thermal quench time proportional
to $S_{wall}.$ 
The $(3,2)$ growth rate is  larger than the  $(2,1).$
The simulations show that the nonlinear behavior is dominated by
the 
$(2,1)$ mode for $q_a \ge  2,$ and the $(3,2)$ for $q_a < 2$.

The TQ is caused by the growth of RWTMs and RWMs.
The linear growth rates indicate the $S_{wall}$ scaling of the TQ time, 
but are not sufficient to obtain $\tau_{TQ}$ quantitatively.
Nonlinear simulations are needed, which are described in the following.

\section{Case $q_a = 2.6$} \label{sec:qa26}

Nonlinear resistive MHD  simulations are carried out with M3D \cite{m3d}
   in which the resistive wall 
\cite{pletzer}  time
was varied. It is found that the MST is unstable to RWTMs and RWMs. 
The modes cause a TQ, on a timescale of order $200ms,$ much longer than the
experimental pulse of $50ms.$ 

The simulations used $16$ poloidal planes, adequate to
resolve low toroidal mode numbers. The simulations 
were dominated by $n = 1,2$ modes.

The simulations were initialized with equilibrium reconstructions in
which $q$ on axis $q_0 = 1,$ and  $q_a = 2.6, 2.0, 1.7$ and $1.5$ at the edge. 

\begin{figure}[h]
\begin{center}
\includegraphics[height=4.8cm]{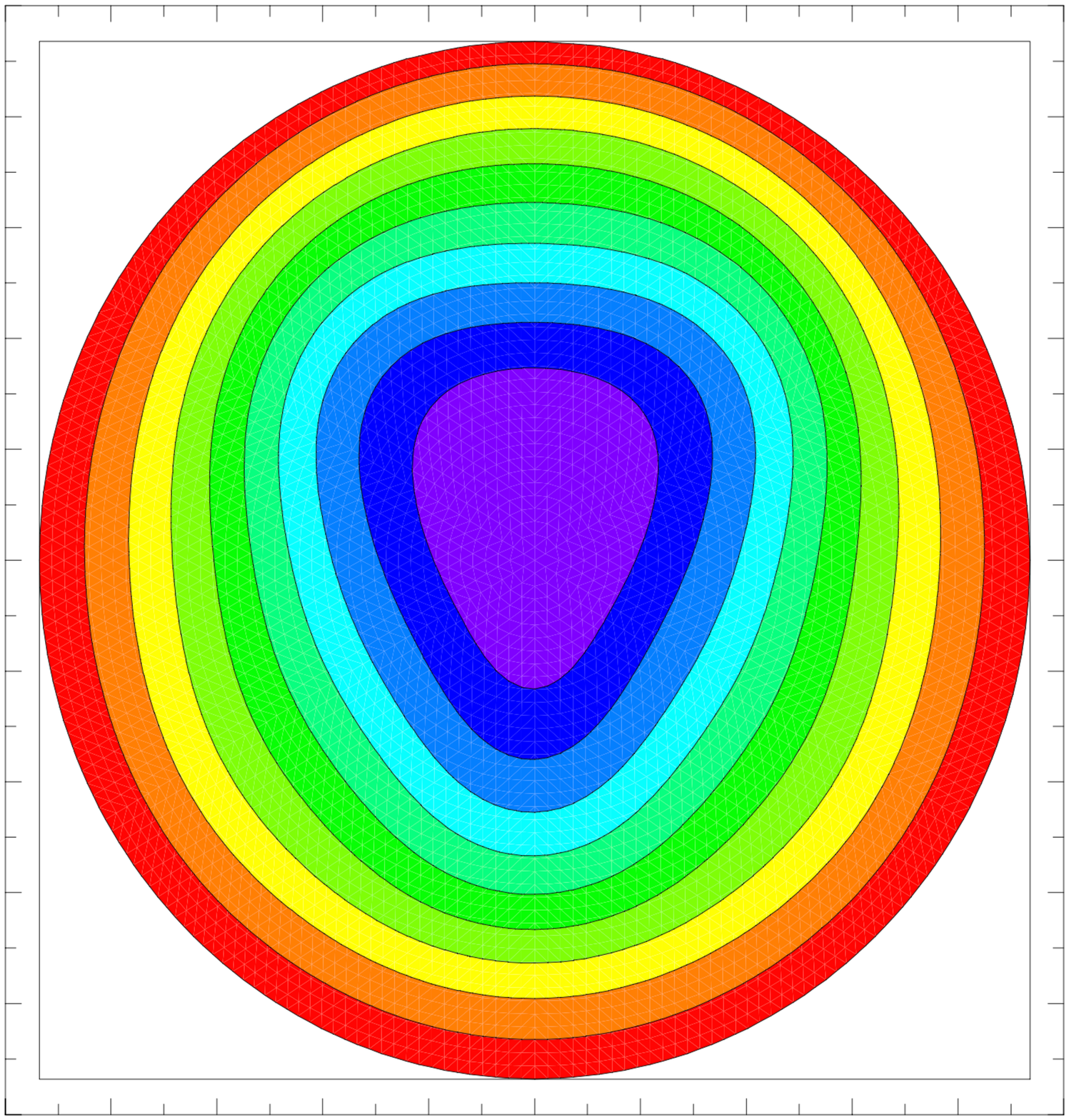}(a)
\includegraphics[height=4.8cm]{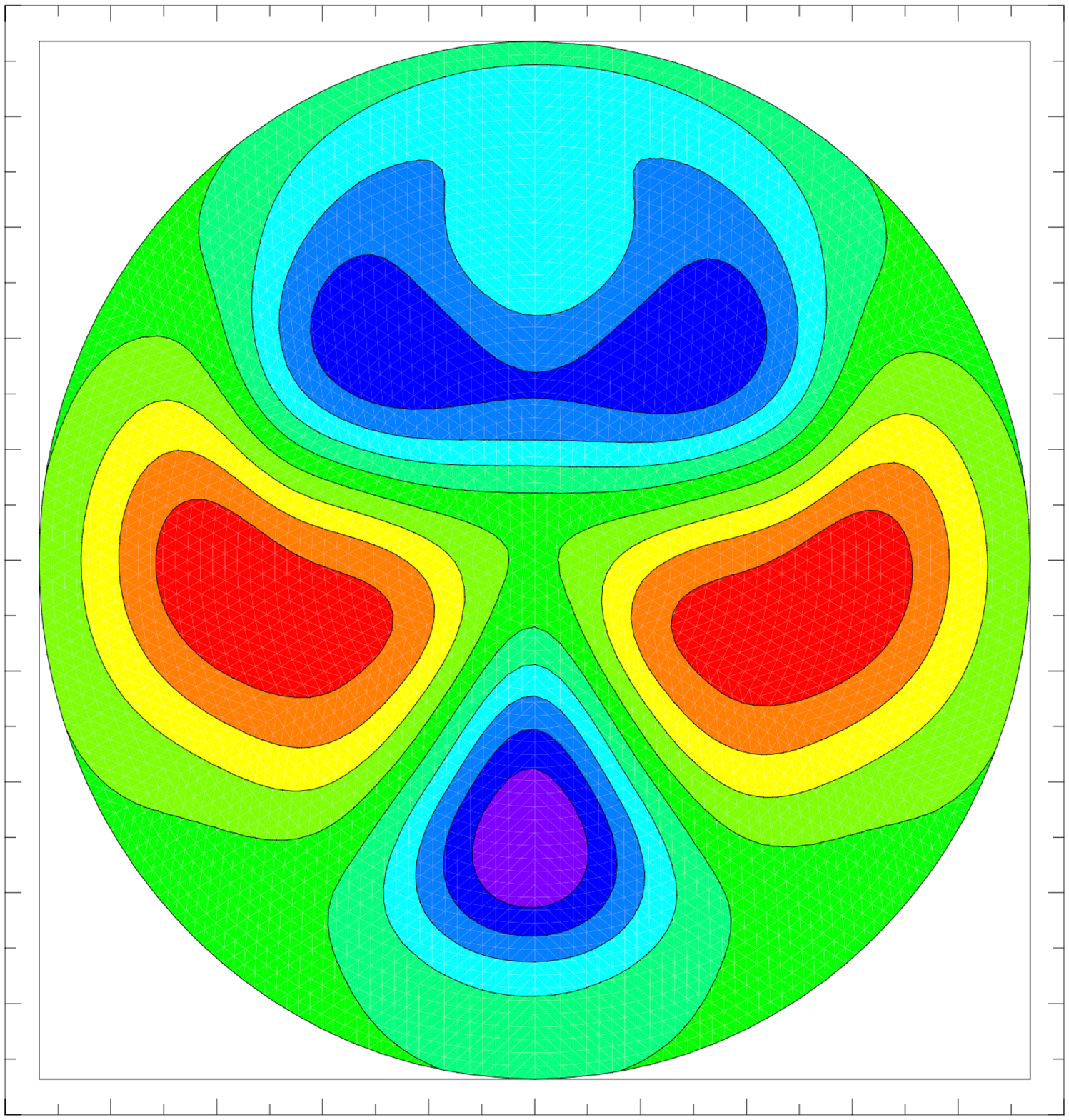}(b)
\includegraphics[height=4.8cm]{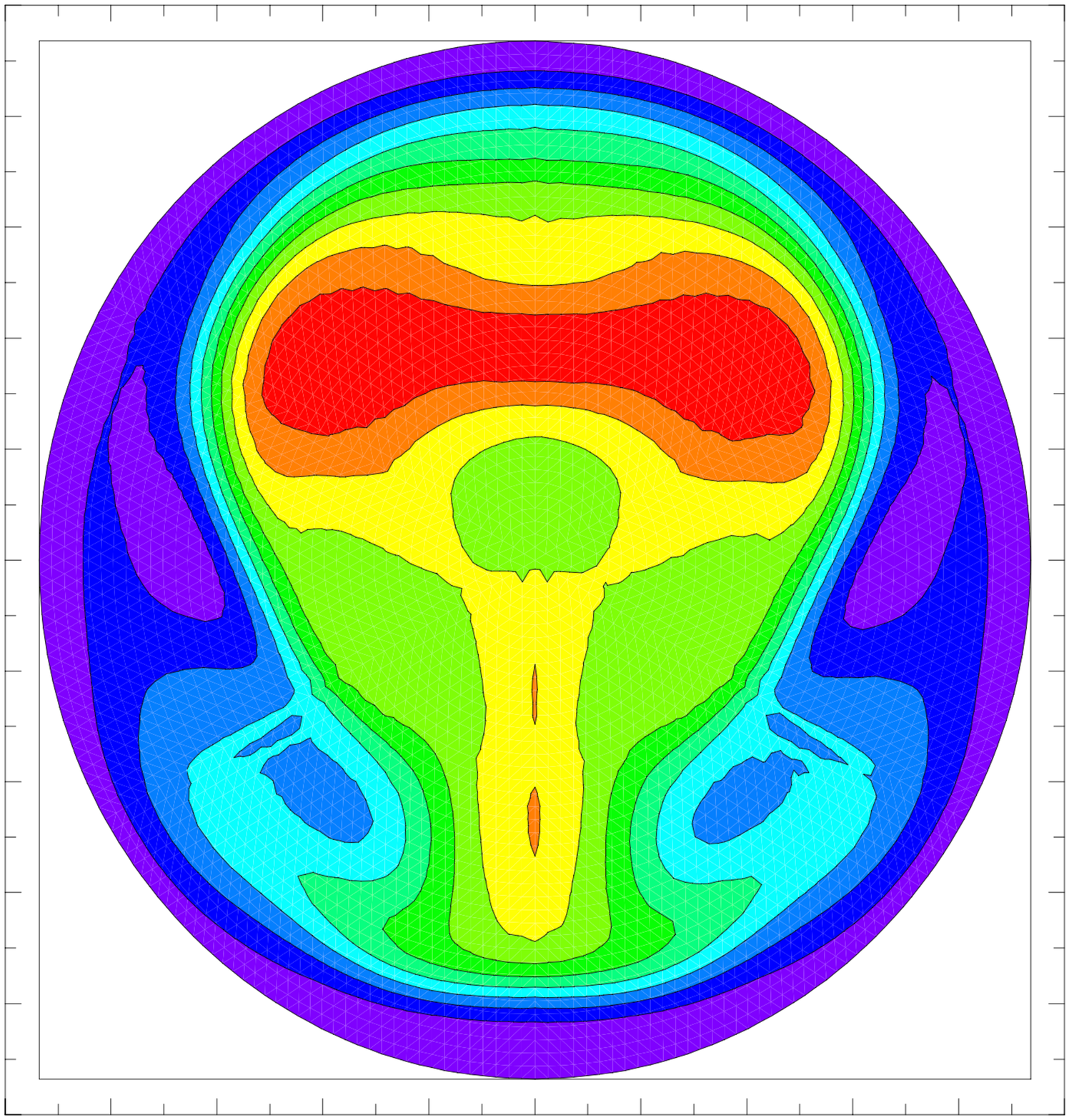}(c)
\end{center}
\caption{\it
(a) contour plot of $\psi$ at time $t = 5300 \tau_A$ for case with
$q_a = 2.6,$ $ S_{wall} = 3.3\times 10^4.$
(b) contour plot of $\tilde{\psi}$ at the same time.
(c) temperature  $T$ at the same time.
} \label{fig:nonlin26}
\end{figure}
\rfig{nonlin26} shows contour plots from a simulation with $q_0,q_a = 1, 2.6.$ It has
$S_{wall}  = 3.3\times 10^4.$ The plots are at time 
$t = 5300\tau_A,$ near the end of the simulation.
\rfig{nonlin26}(a) shows $\psi$, which appears distorted by a $(3,2)$ tearing mode.
The simulation contains  $(3,2)$, and $(2,1)$ modes. 
\rfig{nonlin26}(b) shows the perturbed $\psi,$ where $\tilde{\psi} = \psi - \psi_{avg},$
and $\psi_{avg}$ is the toroidal average. It predominantly has $(2,1)$ structure.
\rfig{nonlin26}(c) is the temperature $T$.

\begin{figure}[h]
\vspace{.5cm}
\begin{center}
\includegraphics[width=7.5cm]{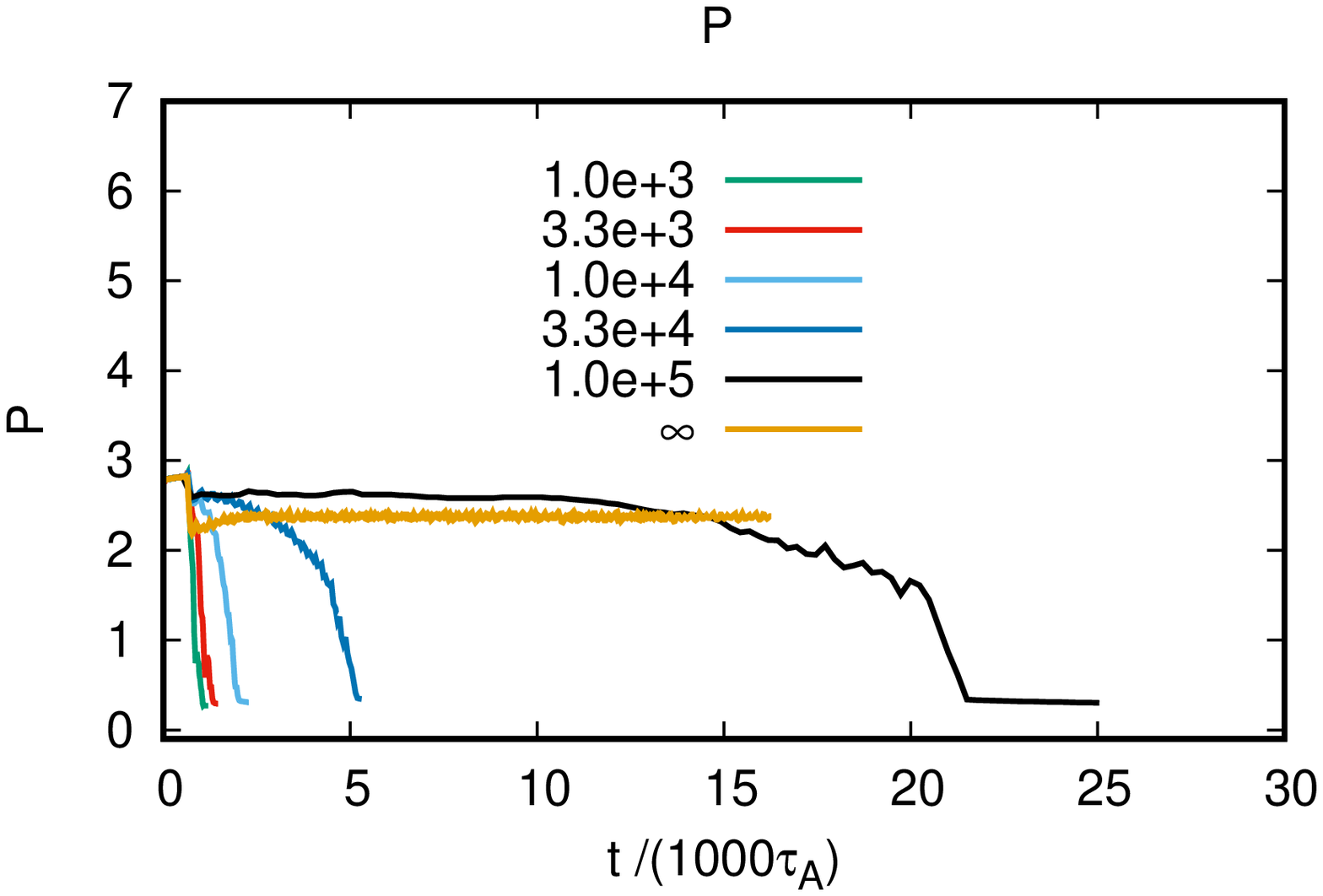}(a)
\includegraphics[width=7.5cm]{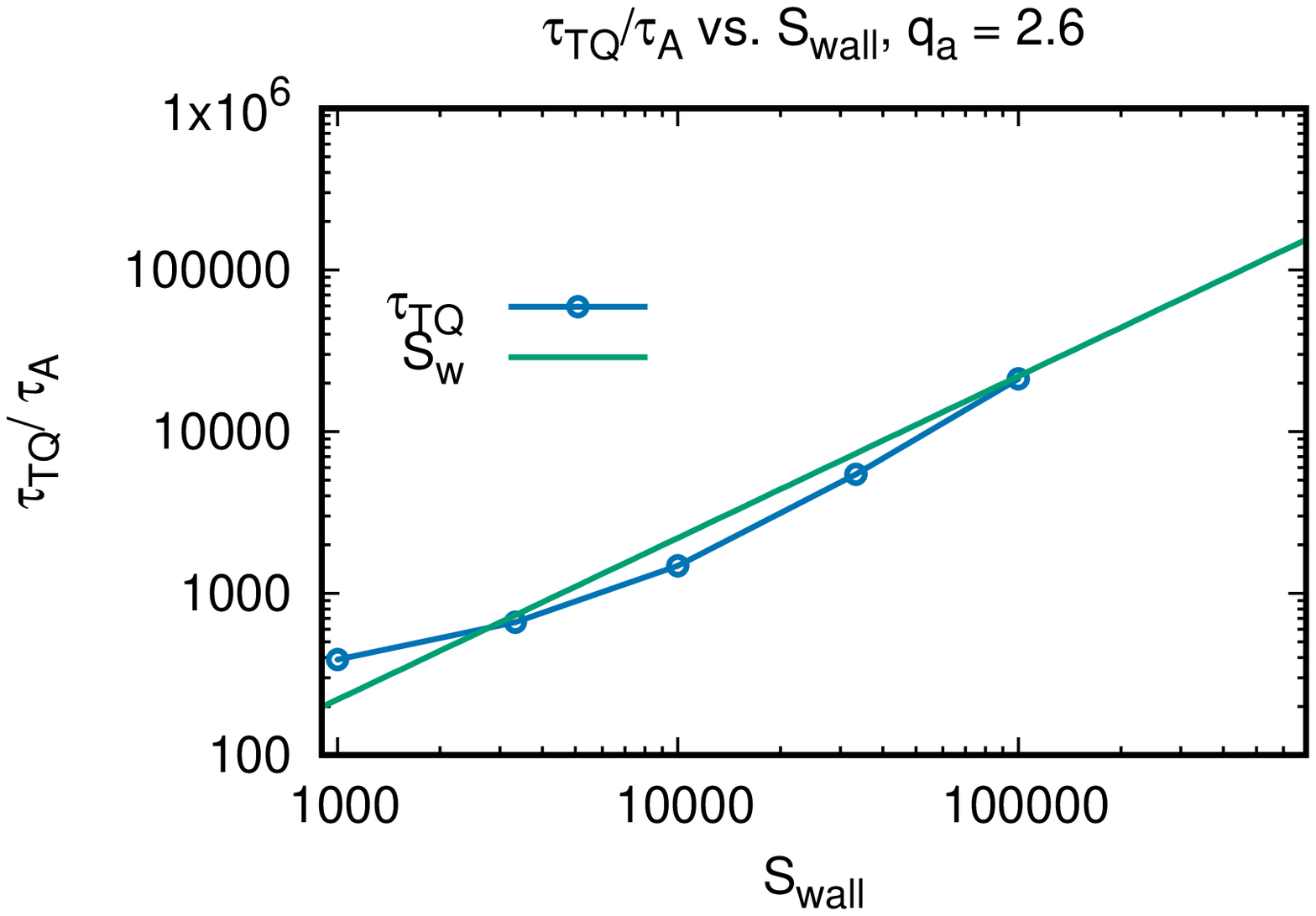}(b)
\end{center}
\vspace{.25cm}
\begin{center}
\includegraphics[width=7.0cm]{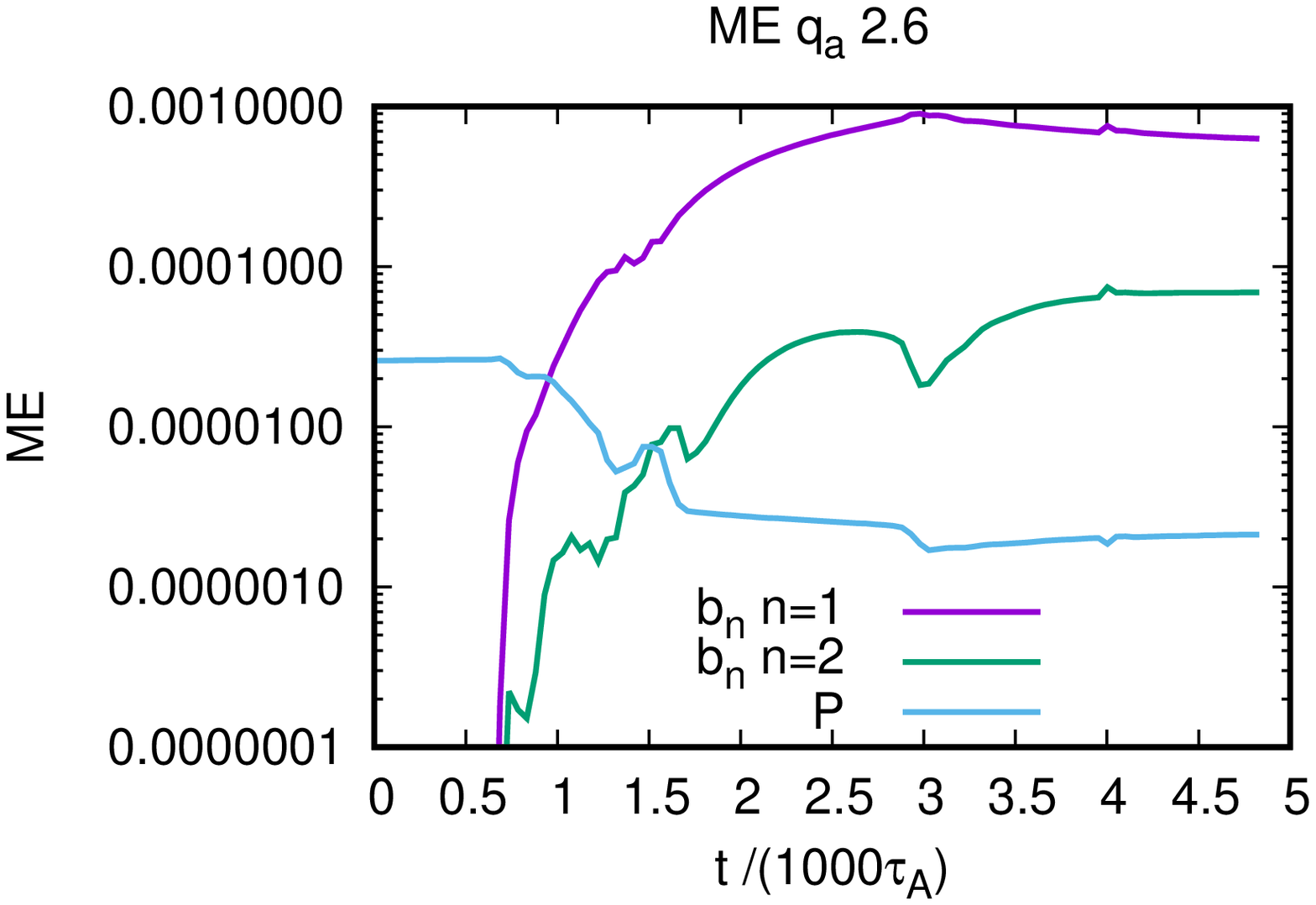}(c)
\end{center}
\vspace{.5cm}
\caption{\it
(a) time history of total pressure $P$ in simulations of MST with $q_a = 2.6.$
The labels denote the value of $S_{wall}.$
The label $\infty$ is for an ideal wall.  In that case there is no TQ.
(b) TQ time $\tau_{TQ}$ measured from the time histories.
The projected TQ time at the experimental $S_{wall} = 7\times 10^5$ is
$\tau_{TQ} \approx 1.5 \times 10^5 \tau_A = 0.18 s.$
(c) Time history of $MW_n,$ the energy of $n = 1, 2$ components of normal magnetic field
at the wall, dominated by a $(2,1)$ mode,  and $P$, 
for $S_{wall} = 10^4.$
}
\label{fig:qa26}
\end{figure}

\rfig{qa26} shows the effect of wall penetration time $\tau_{wall}.$ 
Simulations were done with several values of  $S_{wall}.$ 
 \rfig{qa26}(a) shows the time history of  
the volume integrated  pressure $P$. 
The value of $S_{wall}$ is indicated by labels, 
with $\infty$ indicating an ideal wall.
The TQ time is measured as the 
time difference $(t_{20} - t_{90})/0.7,$ where $t_{90}$ is the time at which the 
temperature is $90\%$ of its peak, and $t_{20}$ is the time when it has 
$20\%$ of its peak value.

The values of $\tau_{TQ}$ as a function  of $S_{wall}$ are collected
in \rfig{qa26}(b). They are fit with $0.22 S_{wall},$ and
projected to the experimental value $S_{wall} = 7\times 10^5,$
$\tau_{TQ} \approx 1.5\times 10^5 \tau_{wall}.$ This is 
$\tau_{TQ} \approx 180 ms,$
the data point for MST plotted in \rfig{ttqs1}.

The magnetic perturbations consist of primarily a $(2,1)$, with a smaller amplitude $(3,2)$ mode.
\rfig{qa26}(c) shows the magnetic energy in $n = 1, 2$ toroidal 
harmonics of the normal component of magnetic field at the wall,
given by  $MW_n = \oint |\bb(n) \cdot \hat{\bf n} / B|^2 dl / L,$  
where $\bb(n) = (2\pi)^{-1} \oint \bb \exp(i n \phi) d \phi,$ $L = \oint dl,$  
$\hat{\bf n}$ is the unit normal to the wall,
and $l$ is the length along the wall for fixed toroidal angle $\phi.$ 
\Rdd{The time dependent increase in the rate of the pressure drop in \rfig{qa26}(a)
is caused by the growth of the amplitude of the magnetic perturbations.} 

\section{Cases $q_a = 2, 1.7, 1.5$} \label{sec:qa2}

It is possible to have $q_a \le 2 $ in MST \cite{mst22}. 
The case $q_a = 2$ 
is at the borderline between RWTM and  RWM. It has a much slower TQ than the case  $q_a = 2.6.$ 

\begin{figure}[h]
\vspace{.5cm} 
\begin{center}
\includegraphics[width=7.5cm]{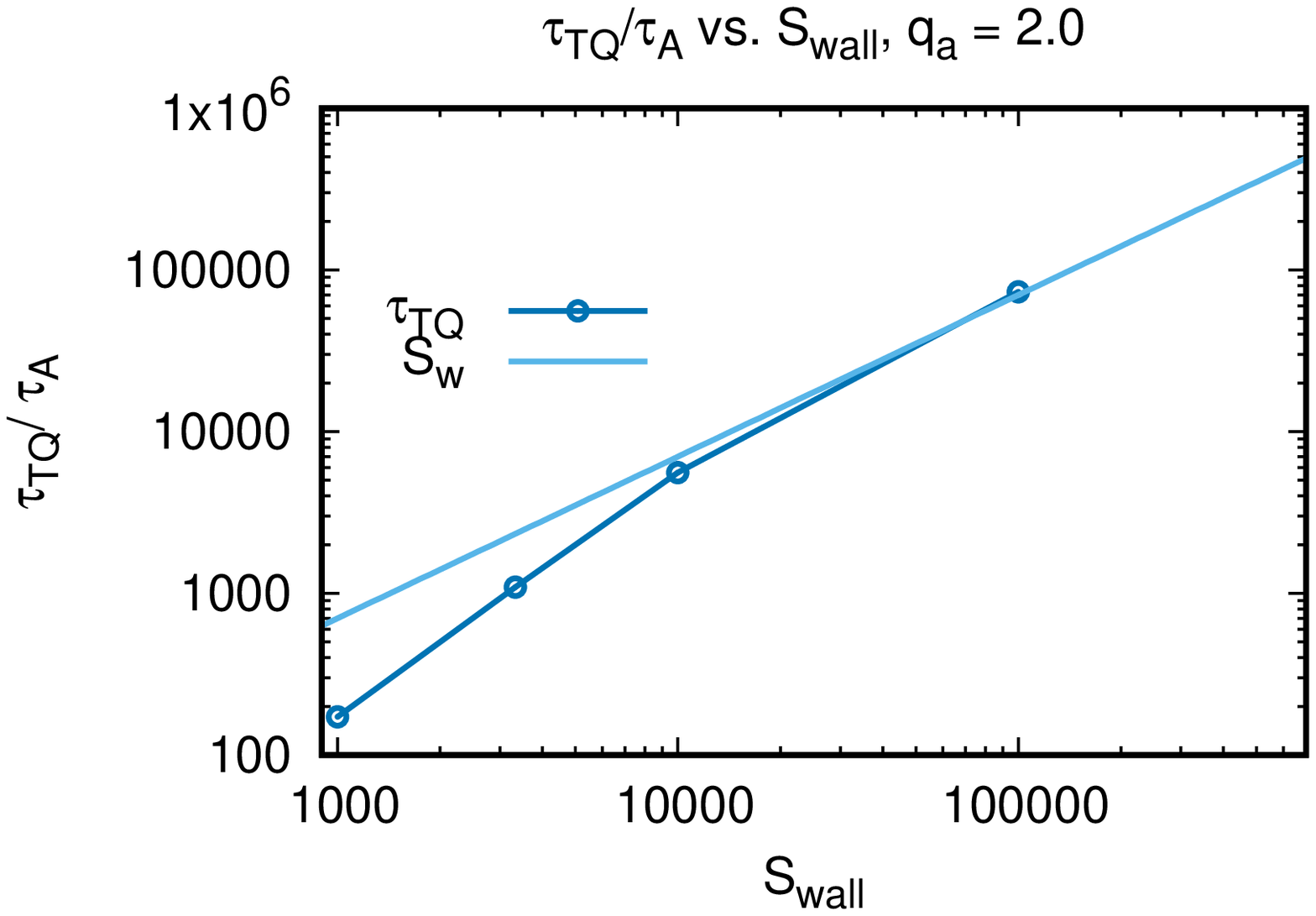}(a)
\includegraphics[width=7.35cm]{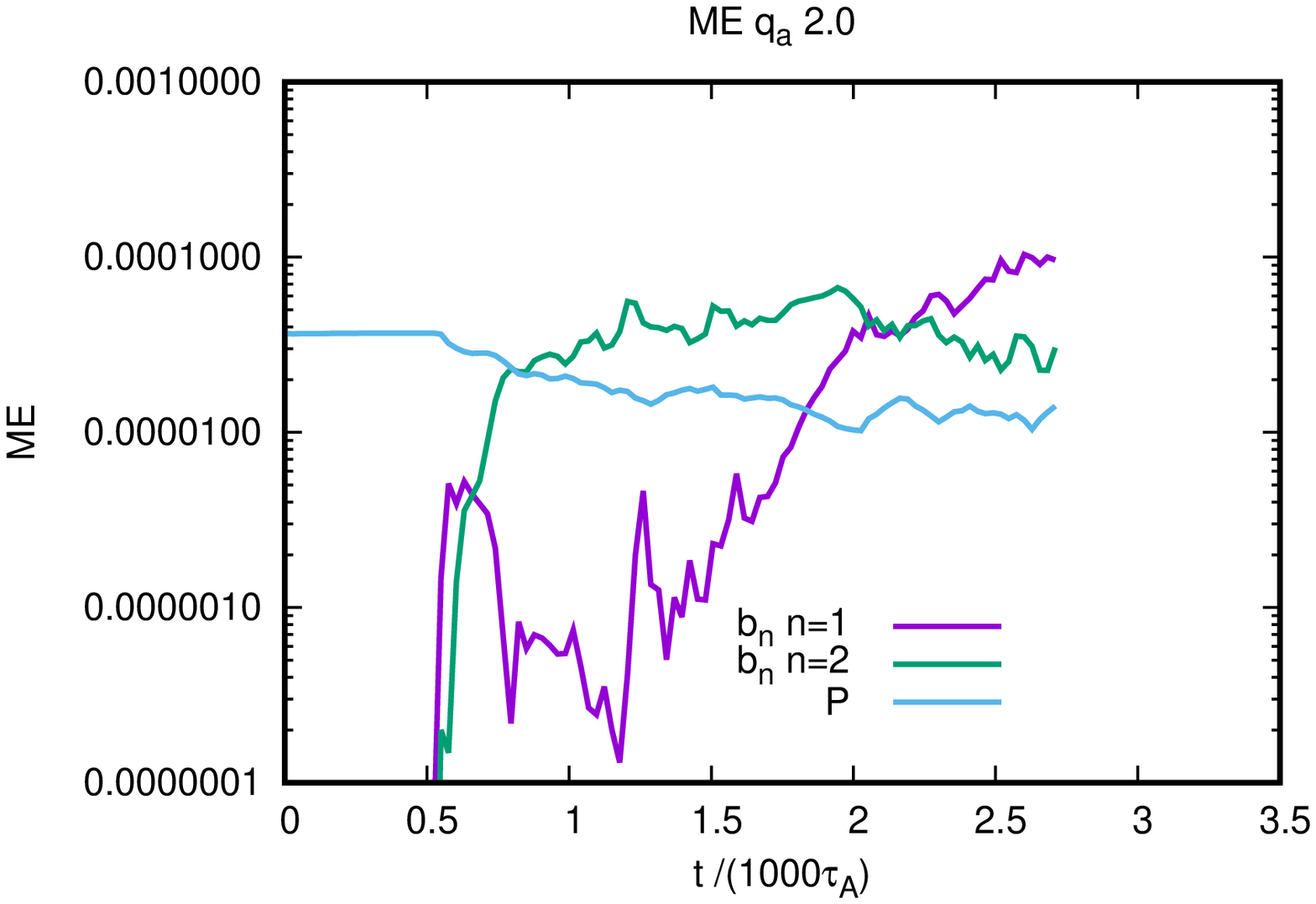} (b)
\end{center}
\vspace{-.5cm}
\caption{\it
(a) plots of $\tau_{TQ}/\tau_A,$ and $0.7 S_{wall}.$ 
The projected TQ time at the experimental $S_{wall}$ is
$\tau_{TQ} \approx 5 \times 10^5 \tau_A = 0.57 s.$
(b) time history of $MW_n,$ wall magnetic normal magnetic energies, and $P$, 
for $S_{wall} = 10^4.$
} \label{fig:q2}
\end{figure}

\rfig{q2}(a) shows the  TQ time as a function of $S_{wall}$ for $q_a = 2.$
Also plotted for comparison is $0.7 S_{wall}.$
Projecting to the experimental $S_{wall}$ gives
$\tau_{TQ} \approx 5\times 10^5 \tau_A$ $\approx 0.57s.$
\rfig{q2}(b) shows $MW_n$ and $P$ for
$S_{wall} = 10^4.$ In this case the $(3,2)$ or  $n= 2$ mode is dominant.  
The $(3,2)$ RWTM has a larger amplitude than the $(2,1)$ RWM. 

\begin{figure}[h]
\vspace{.5cm}
\begin{center}
\includegraphics[width=7.5cm]{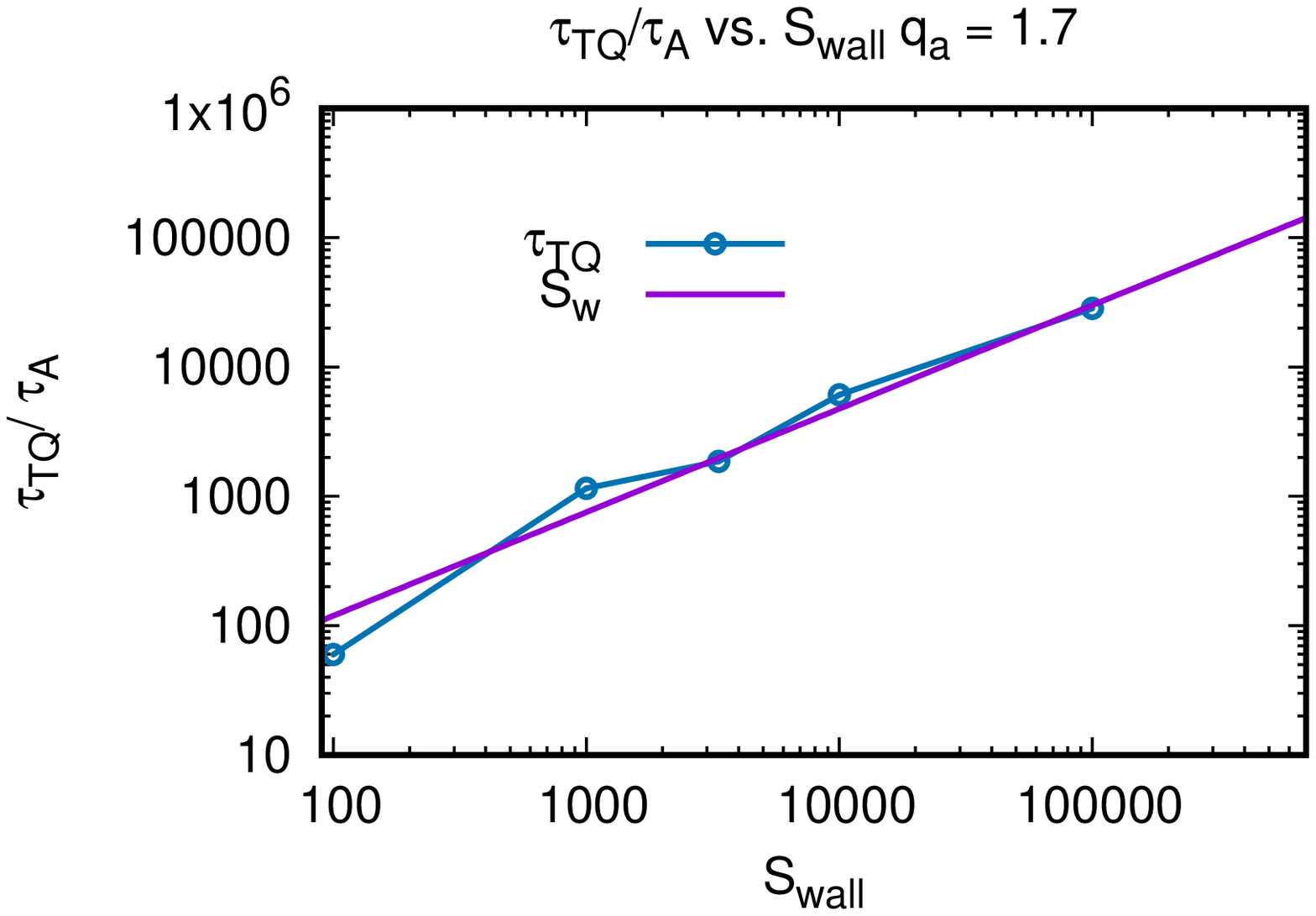}(a)
\includegraphics[width=7.5cm]{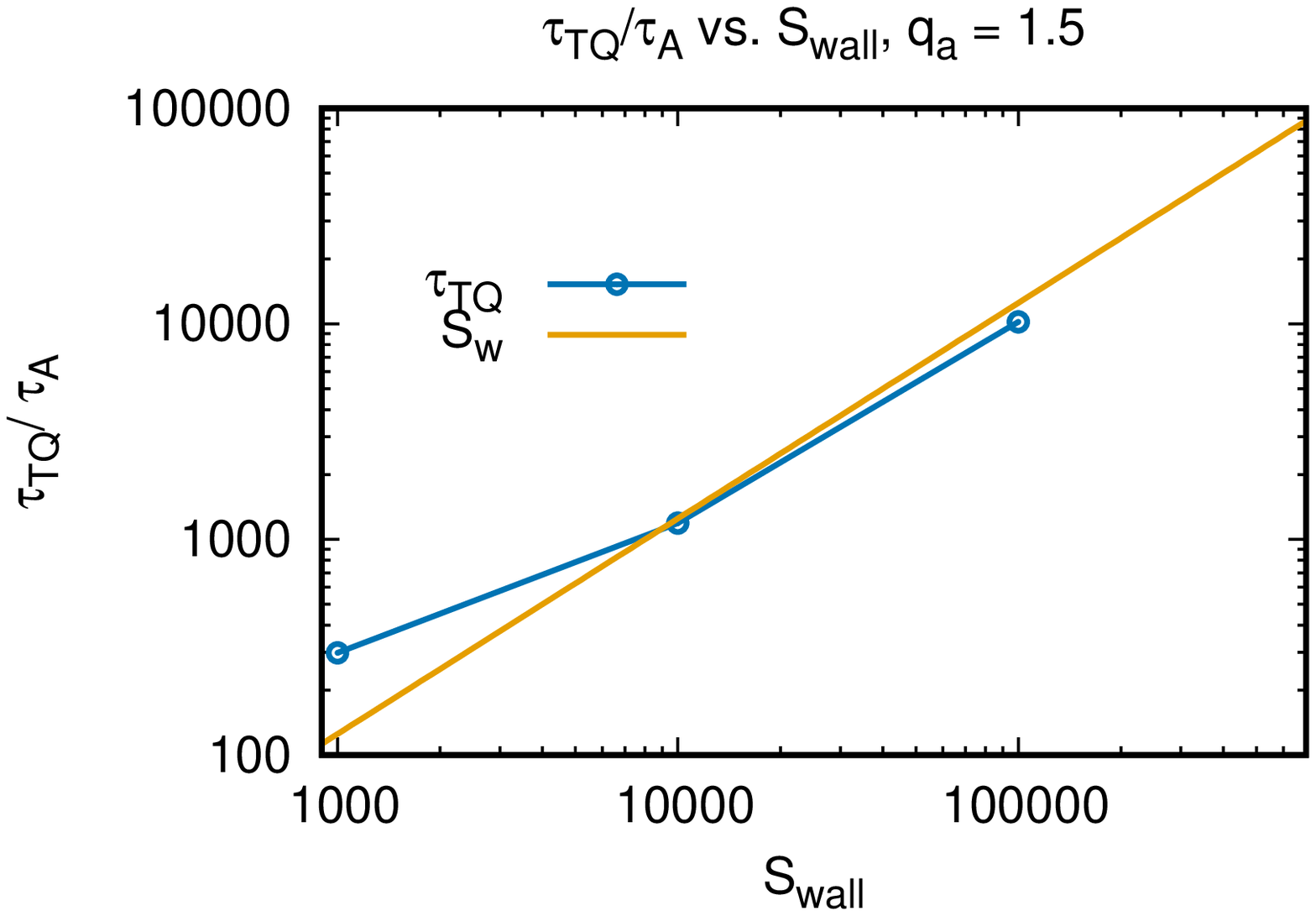}(b)
\end{center}
\vspace{-.5cm}
\caption{\it
(a) TQ time $\tau_{TQ}$ measured from the time histories, for $q_a = 1.7.$
The fit is $\tau_{TQ} \approx 1.7 \tau_{wall},$
which is projected to 
$\tau_{TQ} \approx 1.15\times 10^5 \tau_A,$ $\approx 0.13s.$ 
(b) $\tau_{TQ}$ as a function of $S_{wall},$  for $q_a = 1.5.$ 
Here $\tau_{wall} \approx 0.13 \tau_{wall},$ which is projected to
$\tau_{TQ} \approx 0.9\times 10^5 \tau_A,$ $\approx 0.10s.$} 
\label{fig:tqa17}
\end{figure}

Simulations were also performed for $q_a = 1.7$ and $q_a = 1.5.$
The time history data is summarized in \rfig{tqa17}.
\rfig{tqa17}(a) shows $\tau_{TQ}$ as a function of $S_{wall}$  for $q_a = 1.7.$ 
The data is fit by $\tau_{TQ} \approx 0.16 \tau_{wall},$
which is projected to
$\tau_{TQ} \approx 1.15\times 10^5 \tau_A,$ $\approx 0.13s.$
\rfig{tqa17}(b) shows $\tau_{TQ}$ as a function of $S_{wall}$  for $q_a = 1.5.$
The data is fit by $\tau_{TQ} \approx 0.13 \tau_{wall},$
which is projected to $\tau_{TQ} \approx 0.9\times 10^5 \tau_A,$ $\approx 0.10s.$

\begin{figure}
\vspace{.5cm}
 \begin{center}
\includegraphics[width=7.5cm]{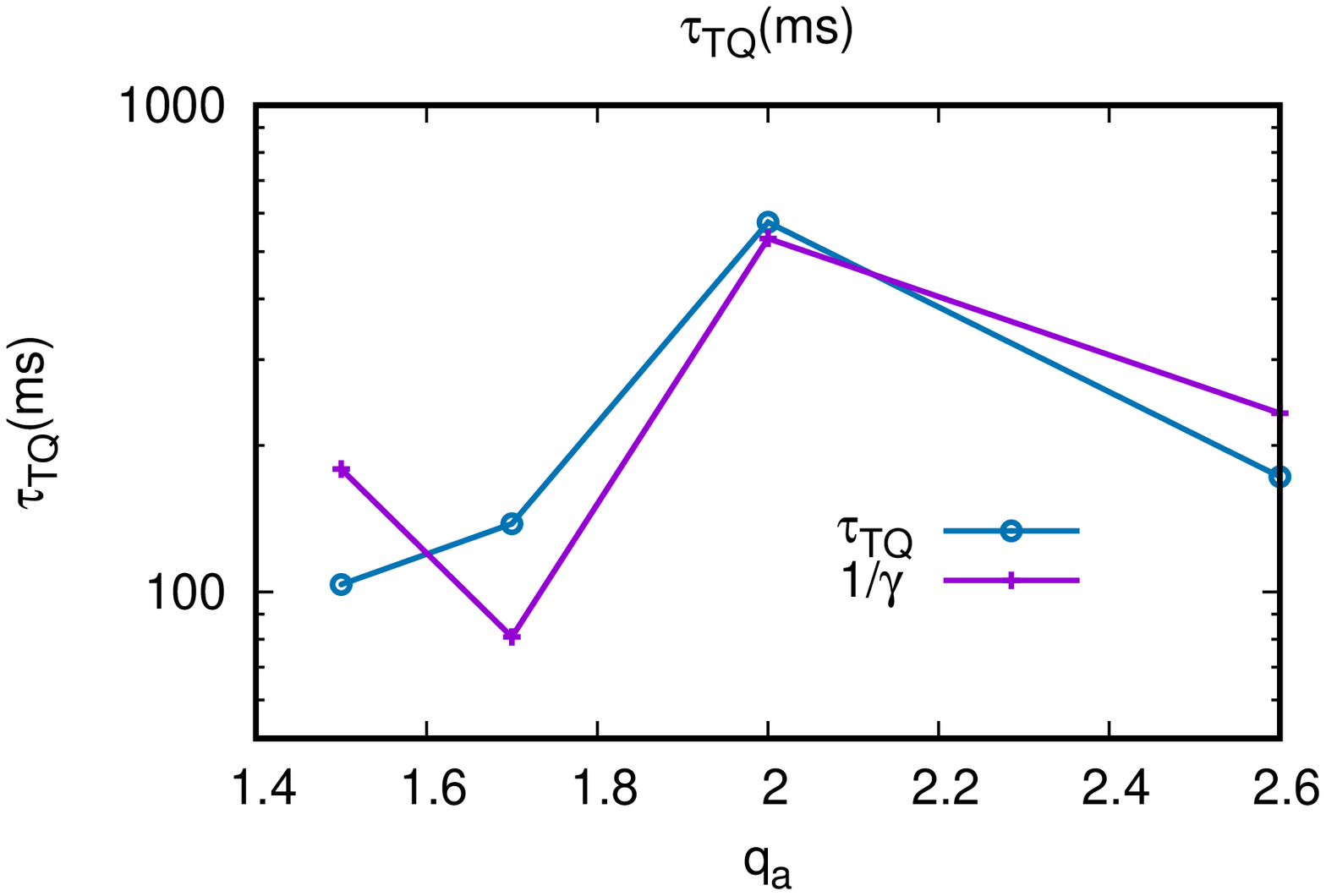}
\end{center}
\caption{\it $\tau_{TQ}$ as a function of $q_a, $ from the simulations,
and $1 / \gamma$ from \rfig{deltai}(a). }
\label{fig:ttqs}
\end{figure}

The TQ time data is summarized in \rfig{ttqs}, showing $\tau_{TQ}$ as
a function of $q_a,$ at the experimental value of $S_{wall}.$ 
Except for $q_a = 2,$ the values of $100ms <\tau_{TQ} < 200ms.$ The
$q_a = 2$ case is much slower, $\tau_{TQ} = 570ms.$ 
Also shown in \rfig{ttqs} is  $1 / \gamma $ 
where $\gamma \tau_{wall}$ is taken from \rfig{deltai}(a), using 
 $\gamma(2,1)$ for $q_a \ge 2,$ and $\gamma(3,2)$ for $q_a < 2.$
\Rdd{The slow TQ at $q_w = 2$ is also seen in the model linear growth times.}
The agreement \Rdd{ of simulations and theory} is remarkable, considering the simplicity of the model
used to obtain  \req{rwm}.

\section{Conclusions} \label{sec:concl}

MST was originally operated as a reversed field pinch, which required a highly
conducting, close fitting wall. 
When run as a tokamak, 
the MST \Rdd{is resistant to  disruptions.} This is consistent with simulations with
$S_{wall}^{-1} = 0,$ which is an ideal wall. When the wall is made
resistive in the simulations, RWTMs become unstable and cause a
thermal quench. The TQ time increases linearly  with $\tau_{wall}$. 
\Rdd{This is characteristic of large $\tau_{wall}$,
$S_{wall} \gg S^{3/5}.$ Asymptotically the RWTM satisfies the RWM dispersion relation, 
except when $\Delta_i \approx 0.$
MST has large $S_{wall}$, as does  ITER when
$S \ll  10^9$ in the edge.} 

Nonlinear simulations examined
four cases, with edge $1.5 \le q_a \le  2.6.$ 
In all cases, the TQ time $\tau_{TQ} > 100 ms,$ compared to the experimental pulse
time $50ms.$ This is shown in \rfig{ttqs}.

The TQ time from the $q_a = 2.6$ case was plotted in \rfig{ttqs1}, since this case more nearly
resembles a standard tokamak.
The implication for other tokamaks is that a more conducting wall slows the
RWTM and mitigates disruptions, especially  in ITER.

{\bf Acknowledgement} 
We thank Jay Anderson for help with the MSTFit code.
This work was supported 
by U.S. DOE under grants DE-SC0020127,  DE-SC0020245, and DE-SC0018266.

\end{document}